\documentclass[prb,superscriptaddress,preprint]{revtex4-1}
\usepackage{SIunits,bm,miller,graphicx,amsmath}

\renewcommand{\vec}[1]{\ensuremath{\boldsymbol{#1}}}
\newcommand{\op}[1]{\ensuremath{\hat{#1}}}
\newcommand{\vecop}[1]{\op{\vec{#1}}}
\newcommand{\ee}{\ensuremath{\mathrm{e}}}
\newcommand{\ii}{\ensuremath{\mathrm{i}}}
\newcommand{\dd}{\ensuremath{\mathrm{d}}}

\begin{document}

\title{Electron vortices in crystals}

\author{Stefan L{\"o}ffler}
\affiliation{Institute of Solid State Physics, Vienna University of Technology, Wiedner~Hauptstra{\ss}e~8-10, 1040 Vienna, Austria}
\email{stefan.loeffler@tuwien.ac.at}
\author{Peter Schattschneider}
\affiliation{Institute of Solid State Physics, Vienna University of Technology, Wiedner~Hauptstra{\ss}e~8-10, 1040 Vienna, Austria}
\affiliation{University Service Center for Transmission Electron Microscopy, Vienna University of Technology, Wiedner Hauptstra{\ss}e 8-10, 1040 Wien, Austria}

\begin{abstract}
The propagation of electron beams carrying angular momentum in crystals is studied using a multislice approach for the model system Fe. It is found that the vortex beam is distorted strongly due to elastic scattering. Consequently, the expectation value of the angular momentum as well as the local vortex components change with the initial position of the vortex and the propagation depth, making numerical simulations indispensable when analyzing experiments.
\end{abstract}

\maketitle

\section{Introduction}

Recently, electron beams carrying topological charge were created successfully in the transmission electron microscope (TEM) \cite{N_v467_i7313_p301,N_v464_i_p737}. The potential applications of these vortex beams (as they were called) seem endless---the speculation is that they will become as versatile a tool as vortices in optics \cite{NP_v3_i5_p305}. In particular, proposed applications range from strongly enhanced energy-loss magnetic chiral dichroism signal over the probing of chiral structures to the manipulation of nanoparticles \cite{NN_v5_i11_p764}.

One essential aspect of electron beams has been overlooked so far, though, which is of paramount importance to real experiments. It is well-known that---due to the strong and long-ranged Coulomb interaction---the probe electrons in a sample undergo strong elastic scattering which depends heavily on aspects such as beam orientation or sample thickness \cite{WilliamsCarter1996,U_v111_i_p1163}. This can be exploited, as in high-resolution TEM (HRTEM) \cite{WilliamsCarter1996}, atom location by channeling-enhanced microanalysis (ALCHEMI) \cite{JoM_v130_i_p147,U_v106_i7_p553}, or energy-loss by channeled electrons (ELCE) \cite{U_v9_i3_p249,S_v218_i4567_p49}. Recently, the new RSTEM technique was used to map the effect of elastic scattering on a focused electron beam with sub-{\AA}ngstrom precision \cite{PRL_v106_i_p160802}.

In other cases, such as energy-loss magnetic chiral dichroism (EMCD) \cite{N_v441_i_p486}, it is usually detrimental to the signal strength \cite{PRB_v75_i21_p214425,U_v111_i_p1163}. In all cases, however, it plays an important role, and the same is to be expected for electron vortex beams \cite{MC2011_vortices}.

\section{Simulations}

In this work, we concentrate on the propagation of electron vortices that have been focused on the specimen, i.e., vortices produced as a convergent beam hitting the specimen. This corresponds to the situation of a holographic or phase plate in the condenser system of a TEM \cite{U_v111_i9-10_p1461,subnm-vortices}. It must be emphasized, however, that the same results also apply to vortex beams that emanate from some point within the sample, e.g., after a chiral excitation as in EMCD.

In the case of a vortex produced by a condenser aperture, we can model the wave in the condenser plane as \cite{U_v111_i9-10_p1461}
\begin{equation}
	\tilde{\psi}(q, \phi) = \ee^{\ii m \phi} \Theta(q_\text{max} - q),
\end{equation}
where $m = 1$ is the topological charge, $\phi$ is the polar angle in the condenser plane, $q$ is the radial vector in the condenser plane, $q_\text{max}$ is the radius of the aperture (which defines the maximum convergence angle), and $\Theta(x)$ is the Heaviside function. In the focused case (ignoring aberrations), the wave function $\psi$ hitting the sample is given by the Fourier transformation of $\tilde{\psi}$, which corresponds to the Hankel transformation in polar coordinates \cite{subnm-vortices}:
\begin{equation}
\begin{aligned}
	\psi(r, \varphi) 
	&= \frac{\ii \ee^{\ii \varphi}}{2\pi} \int_0^{q_\text{max}} J_1(q r) q \dd q \\
	&= \frac{\ii q_\text{max} \ee^{\ii \varphi}}{4r}\left( J_1(q_\text{max} r) H_0(q_\text{max} r) - J_0(q_\text{max} r) H_1(q_\text{max} r) \right).
\end{aligned}
\end{equation}
Here, we used the Bessel functions of first kind ($J_m$) and the Struve functions ($H_m$). 

\begin{figure}
	\centering
	\includegraphics{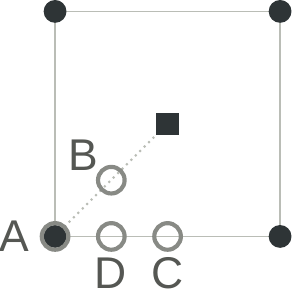}
	\caption{Projection of an Fe crystal in \hkl[0 0 1] direction. Atoms in different layers are drawn as disks and squares, respectively. The positions mentioned in the text are marked by capital letters. The edge length of the square is the lattice constant of Fe, namely \unit{2.859}{\angstrom}.}
	\label{fig:zone-axis}
\end{figure}

As model system, we have chosen Fe. On the one hand, it has a simple crystal structure (body centered cubic), on the other hand it is an easily accessible sample material and of general interest as a ferromagnetic model system\footnote{The effect of the internal magnetic field on the probe was ignored in this work, however, as is common procedure in elastic calculations.}. In particular, we investigated the \hkl[0 0 1] zone axis which is a high-symmetry zone axis as can be seen from fig.~\ref{fig:zone-axis}.

For the simulations, the focused vortex beam was positioned on four different positions on the crystal (see Fig.~\ref{fig:zone-axis}): on an atomic column (A), between the two closest atomic columns (B), as far away from atomic columns as possible (C), and in another, non-symmetric place (D). In all cases, a thickness of \unit{200}{\angstrom} and a spherical aberration coefficient $C_S = 0$ were assumed. Furthermore, an energy of \unit{200}{\kilo\electronvolt}, an angular momentum of $\hbar$ and a convergence semi-angle of \unit{15}{\milli\rad} (corresponding to a waist radius of \unit{\approx 0.67}{\angstrom}) were used for the incident electron beam.

\begin{figure*}
	\centering
	\includegraphics{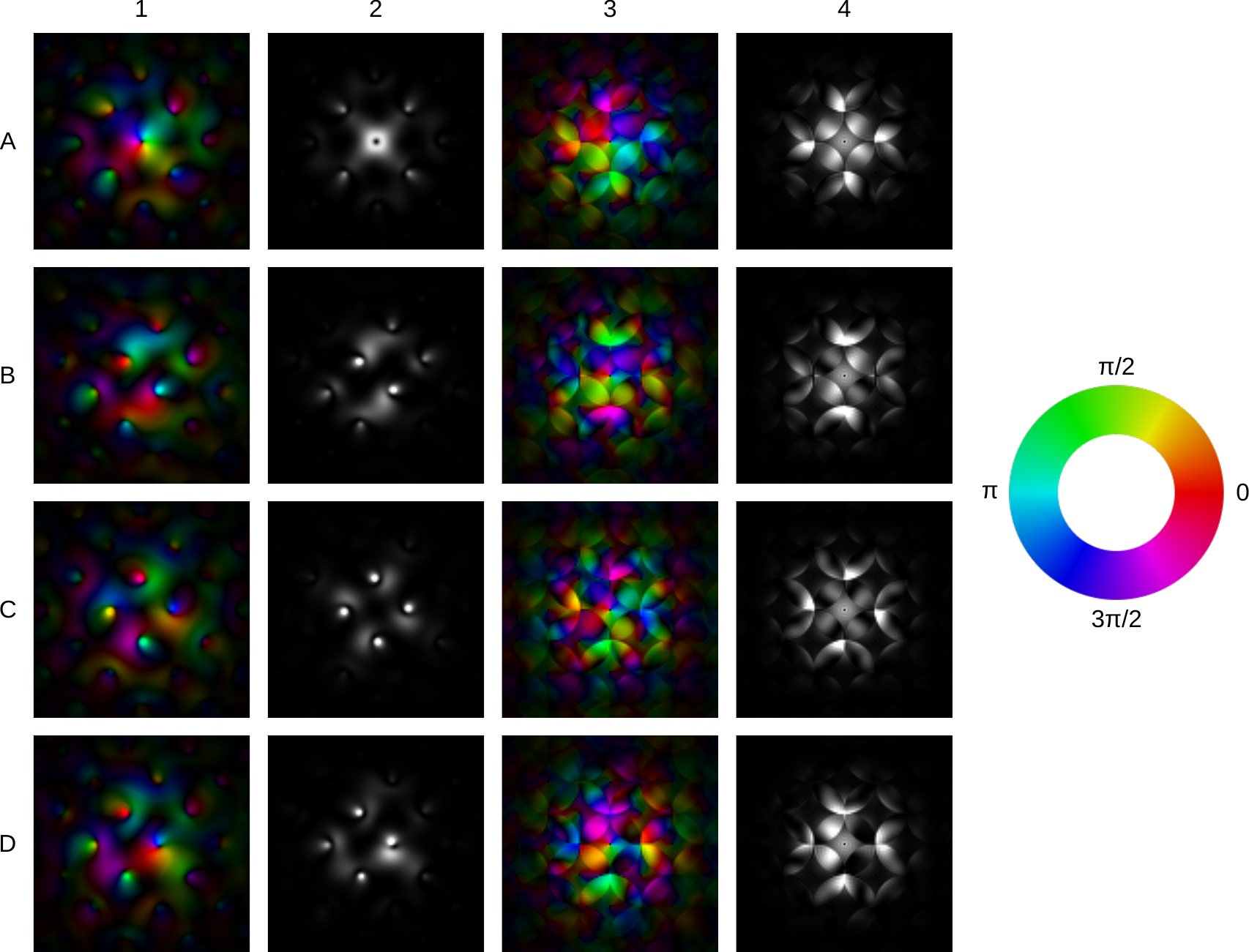}
	\caption{Wave functions at the exit plane of the \unit{20}{\nano\meter} thick Fe sample for the positions A, B, C, and D as defined in the text and in fig.~\ref{fig:zone-axis}. Column 1 shows the real-space amplitudes (brightness) and phases (hue, as specified in the color wheel on the right), column 2 shows the corresponding intensities, column 3 shows the reciprocal-space amplitudes and phases, and column 4 shows the diffraction intensities. The 1 and 2 show a $(\unit{10}{\angstrom})^2$ subset of the simulated data, while the columns 3 and 4 show a $(\unit{2.45}{\per\angstrom})^2$ subset.
	}
	\label{fig:images}
\end{figure*}

This electron beam was subsequently propagated through the crystal using the multislice approach \cite{Kirkland1998,Zauner2010,U_v96_i3-4_p343}. Fig.~\ref{fig:images} shows the propagated wave functions at the exit plane. From them, it is obvious that (a) the wave functions do no longer show the typical vortex phase and amplitude structure \cite{U_v111_i9-10_p1461} in general, (b) the structure---and in particular the phase distribution---depends critically on the initial position of the focused vortex with respect to the crystal, and (c) that we generally see curved structures as one would expect for vortices, but which are absent in normal STEM \cite{U_v96_i3-4_p343}.

In addition to the curved features between the atomic columns, the simulations shown in fig.~\ref{fig:images} also reproduce the channeling and dechanneling effects usually seen with focused electron beams \cite{U_v96_i3-4_p343}. As a consequence, one sees significant intensity not only close to the initially illuminated atomic column, but also on the nearest and next-nearest neighboring columns. Moreover, the central vortex phase structure is relatively well-preserved in-between the atomic columns, even after a \unit{200}{\angstrom} thick sample. In close proximity of the atomic columns it is heavily distorted, however, which also leads to very different local phases at and around neighboring atoms.

It must be noted here that all the images in fig.~\ref{fig:images} display the wave functions directly at the exit plane of the sample. If one wants to observe such patterns experimentally in a microscope, one has to use some post-specimen lens system to transfer those wave functions to the screen. These lens systems are usually not ideal and hence can introduce artifacts or obscure some details. Since the important object for all interactions in the sample is the wave function, however, it is more appropriate to show this directly, rather than the image one would get on a camera.

Recently, the feasibility of recording images similar to those shown in fig.~\ref{fig:images} was also demonstrated experimentally using the new RSTEM technique \cite{PRL_v106_i_p160802}. In combination with a next-generation, double-corrected TEM, this technique can be used to investigate the real-space image of a focused electron beam with sub-{\AA}ngstrom resolution after it has traversed the crystal.

\section{Angular momentum}

One of the most important characteristics of a vortex beam is angular momentum. As helical wave functions are eigenstates of the $\op{L}_z$ operator \cite{U_v111_i9-10_p1461}, they have a well-defined, integer angular momentum quantum number $m$ which is a constant of motion in free space. In fact, it is precisely this quantum number that defines the ``vorticity'' of the beam. But how does it behave inside the sample?

\begin{figure}
	\includegraphics{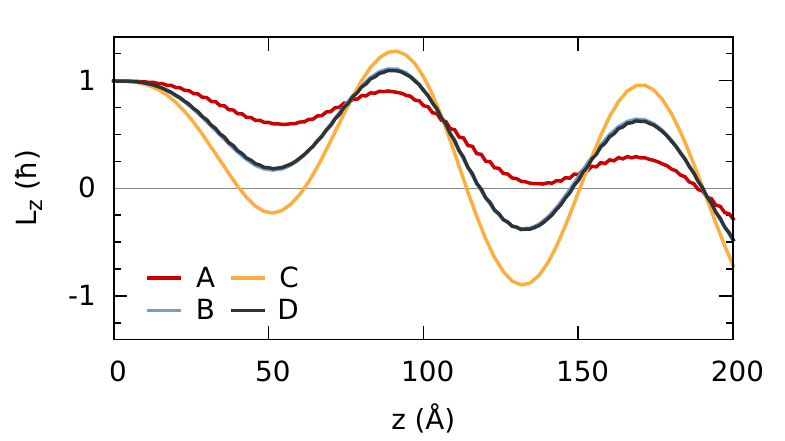}
	\caption{Expectation value $\langle\op{L}_z\rangle$ in Fe \hkl[0 0 1] for the four positions given in the text and in fig.~\ref{fig:zone-axis} as a function of penetration depth $z$. The curves for the positions B and D overlap.}
	\label{fig:lz}
\end{figure}

Fig.~\ref{fig:lz} shows the evolution of the expectation value $\langle \op{L}_z \rangle$ of the angular momentum as a function of depth $z$ in the crystal. It is clearly obvious that (a) it changes significantly as the probe propagates through the crystal, (b) that it is generally not an integer and can even surpass the initial value of $\hbar$, and (c) that it is position-dependent.

All of this can be explained by looking at the properties of $\op{L}_z$. While this operator commutes with spherical operators such as the free-space Hamiltonian, it does not generally commute with the crystal Hamiltonian in which the spherical symmetry is broken by the crystal potential. Thus, $\langle \op{L}_z \rangle$ is constant in free space, but can and will change inside a crystal. This also implies that the probe inside the crystal will evolve into a (coherent) superposition of different eigenstates of $\op{L_z}$ over time, accounting for the non-integer expectation values.

Physically, this can be understood as an exchange of angular momentum between the probe beam and the crystal. Because of the very different masses of probe electron and sample as well as the oscillatory nature\footnote{Interestingly, the period length of $\approx\unit{8}{\nano\meter}$ of the \protect{$\langle \op{L}_z \rangle$} oscillations is of the same order of magnitude as the $\approx\unit{10}{\nano\meter}$ period length of the pendell{\"o}sung.} of $\langle \op{L}_z \rangle$ in fig.~\ref{fig:lz}, the change of the angular momentum of the sample usually is not measurable. For nanoparticles of suitable sizes, however, this could produce appreciable angular velocities.

From this physical interpretation, the position dependence is understandable as well. If the ({\AA}ngstrom sized) probe is positioned directly on an atom column, it usually channels down along that column, whereas a similar probe positioned between columns will first spread until it hits several columns at once and will subsequently channel down primarily along those \cite{U_v96_i3-4_p343}. Therefore, the amount the focused probe beam interacts with the sample obviously depends on its position, as is readily visible in fig.~\ref{fig:lz}.

\section{Vortex decomposition}

\begin{figure}
	\includegraphics{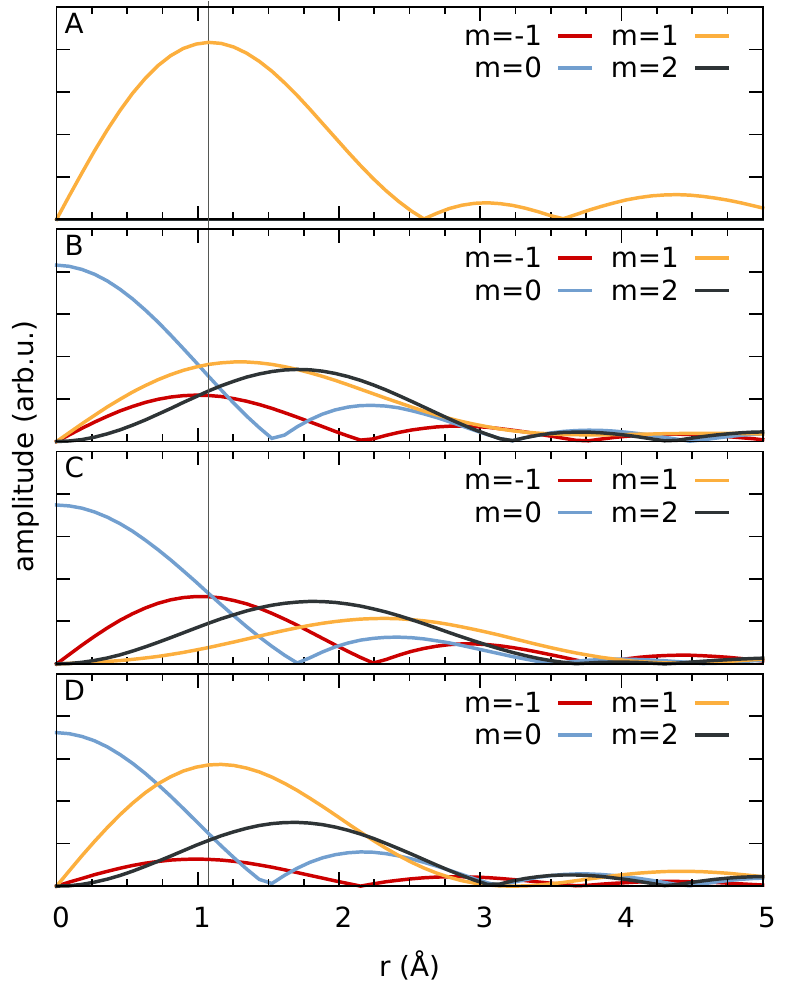}
	\caption{Expansion of an $m=1$ vortex incident on position A for the positions A--D given in the text. Only the most important components $ml=-1, 0, 1, 2$ are plotted. The scale is the same in all subfigures. The vertical line marks the nominal beam waist radius that was used to calculate the depth plots in fig.~\ref{fig:vortex_depth_A}.}
	\label{fig:vortex_0000_A}
\end{figure}

In the previous section, we have seen that inside the sample, a pure vortex state evolves into a coherent superposition of such states. In fact, the occurrence of different vortex components is not an effect unique to a crystal---even in free space, a vortex wave can be described by a single $\exp(\ii m \varphi)$ function (with polar angle $\varphi$) only in a single cylindrical coordinate system, namely the one for which the $z$ axis coincides with the propagation axis of the phase singularity in the center of the vortex. If the vortex is displaced relative to the origin of the coordinate system, again several components come into play, as can readily be seen from fig.~\ref{fig:vortex_0000_A}. Only if we describe the $m=1$ vortex centered on position A around that very position do we retrieve exclusively an $m=1$ component. In all other cases, additional components are mixed in, including $m=0$ Airy-like components and even opposite $m=-1$ components.

In free space, all these components add up to a single (displaced) vortex wave which has a well-defined and---most importantly---constant angular momentum\footnote{It can readily be seen that---similar to the parallel axis theorem in classical mechanics---\protect{$\langle \op{L}_z \rangle = m \hbar + \ii \hbar \vec{r}_0 \times \langle \vecop{p} \rangle$} under a displacement $\vec{r}_0$.}. In a crystal, however, this is no longer the case, and consequently the relative weights of the different components change with the penetration depth in the crystal, just as $\langle\op{L}_z\rangle$ changes.

\begin{figure}
	\includegraphics{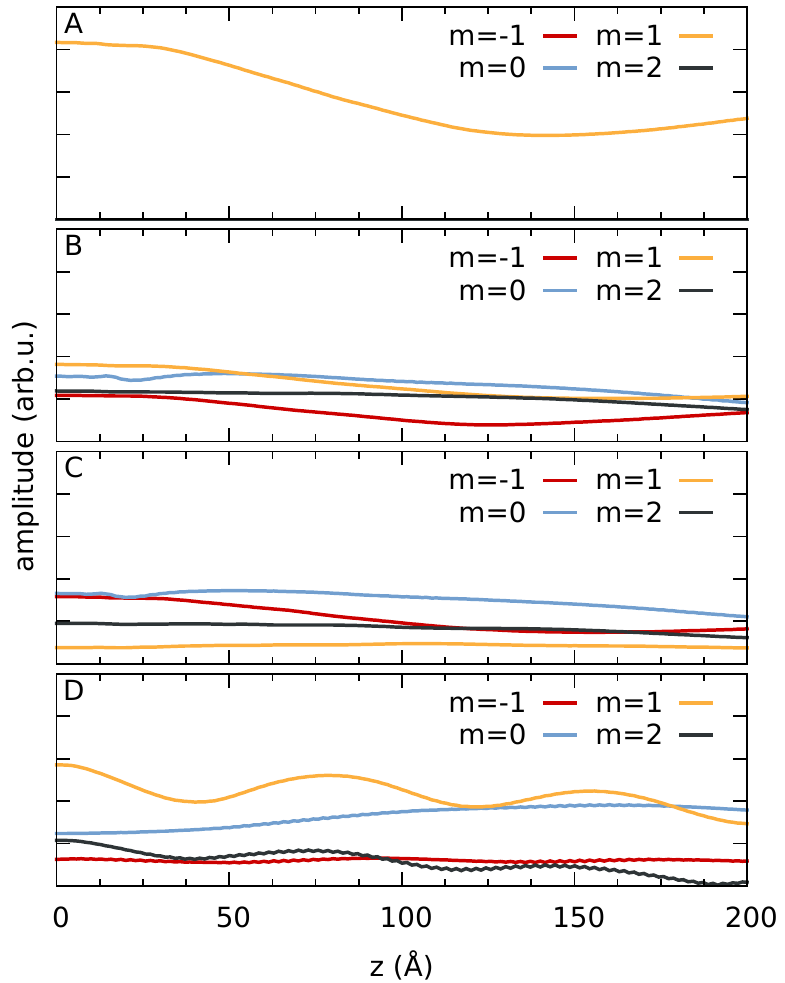}
	\caption{Dependence of the local amplitudes of the most important vortex components $m = -1, 0, 1, 2$  for vortices positioned at the positions A--D given in the text on the depth $z$ in the crystal. The local amplitudes were calculated at a radius of $r = \unit{1.08}{\angstrom}$ (which corresponds to the nominal beam waste radius) around the atom at position A. The scale in all subfigures is the same as in fig.~\ref{fig:vortex_0000_A}.}
	\label{fig:vortex_depth_A}
\end{figure}

Fig.~\ref{fig:vortex_depth_A} shows how the most important local amplitude components around the atom at position A change with depth, dependent on the vortex position. For this, the vortex was propagated with an initial position A--D. At each depth, the wave function was decomposed in vortex components around the atom A, and the amplitude at the radius $r = \unit{1.08}{\angstrom}$ (which corresponds to the nominal beam waist radius) was extracted. As such, $z=0$ in fig.~\ref{fig:vortex_depth_A} corresponds to $r = \unit{1.08}{\angstrom}$ in fig.~\ref{fig:vortex_0000_A}.

It is readily visible from fig.~\ref{fig:vortex_depth_A} that the local vortex amplitudes change with thickness, but the precise behavior depends crucially on the initial vortex position and the vortex order of interest. The oscillatory behavior seen best in the curves for A~/~$m=1$, D~/~$m=1$, and D~/~$m=2$ is reminiscent of the pendell{\"o}sung of conventional Bloch waves, although an analytical equation for its description is unfeasible because each vortex contains infinitely many Bloch wave components.

\section{Conclusion}

The calculations shown in the previous sections show that one has to be careful when predicting or interpreting the effects of vortices if they pass through matter. In particular, statements one usually takes for granted when dealing with vortices---such as the conservation of topological charge or equivalently angular momentum---are only valid in free space. Inside a sample, electron vortex beams can be heavily distorted, which will change the components present in a local environment around the atom. As a consequence, one can have $m=0$ and $m=2$ beams in the vicinity of a scattering atom, even if the original vortex had $m=1$.

This creates new challenges for the use of vortices in, e.g., EMCD, but at the same time provides new exciting possibilities. These range from the conversion of vortex beams from one topological charge to another to the transfer of angular momentum to nanoparticles.

\begin{acknowledgements}
S.L. and P.S. acknowledge the support of the Austrian Science Fund (FWF) under grant number I543-N20.
\end{acknowledgements}


%

\end{document}